\documentclass{article}

\usepackage{microtype}
\usepackage{graphicx}
\usepackage{epstopdf}
\usepackage{subfigure}
\usepackage{booktabs} % for professional tables
\usepackage{multirow}

\usepackage{hyperref}

\usepackage[accepted]{icml2024}

\usepackage{amsmath}
\usepackage{amssymb}
\usepackage{mathtools}
\usepackage{amsthm}

\usepackage[capitalize,noabbrev]{cleveref}

\theoremstyle{plain}

\theoremstyle{definition}

\theoremstyle{remark}

\usepackage[textsize=tiny]{todonotes}

\icmltitlerunning{AdaNovo: Adaptive \emph{De Novo} Peptide Sequencing with 
Conditional Mutual Information}

\begin{document}

\twocolumn[
\icmltitle{AdaNovo: Adaptive \emph{De Novo} Peptide Sequencing with \texorpdfstring{\\} CConditional Mutual Information}

% It is OKAY to include author information, even for blind
% submissions: the style file will automatically remove it for you
% unless you've provided the [accepted] option to the icml2024
% package.

% List of affiliations: The first argument should be a (short)
% identifier you will use later to specify author affiliations
% Academic affiliations should list Department, University, City, Region, Country
% Industry affiliations should list Company, City, Region, Country

% You can specify symbols, otherwise they are numbered in order.
% Ideally, you should not use this facility. Affiliations will be numbered
% in order of appearance and this is the preferred way.
\icmlsetsymbol{equal}{*}

\begin{icmlauthorlist}
\icmlauthor{Jun Xia}{equal,wl}
\icmlauthor{Shaorong Chen}{equal,wl}
\icmlauthor{Jingbo Zhou}{equal,wl}
\icmlauthor{Tianze Ling}{thu}
\icmlauthor{Wenjie Du}{wl}
\icmlauthor{Sizhe Liu}{usc}
\icmlauthor{Stan Z. Li}{wl}
\icmlcorrespondingauthor{Stan Z. Li}{stan.zq.li@westlake.edu.cn}
% 
%\icmlauthor{}{sch}
\end{icmlauthorlist}

\icmlaffiliation{wl}{School of Engineering, Westlake University}
\icmlaffiliation{thu}{Tsinghua Univerisity}
\icmlaffiliation{usc}{University of Southern California}

% \icmlcorrespondingauthor{Firstname2 Lastname2}{first2.last2@www.uk}

% You may provide any keywords that you
% find helpful for describing your paper; these are used to populate
% the "keywords" metadata in the PDF but will not be shown in the document
\icmlkeywords{Machine Learning, ICML}
\vskip 0.3in
]

\printAffiliationsAndNotice{\icmlEqualContribution} % otherwise use the standard text.

\begin{abstract}
% Tandem mass spectrometry has become the driving force behind the expansion of proteomics by serving as a potent method for analyzing the protein composition of biological samples. Various deep learning methods have been developed to identify the amino acids sequence (i.e., the peptide) that is responsible for generating the observed spectrum. However, two main challenges hinder the further development of these \emph{de novo} peptide sequencing methods. Firstly, we observe that Post-Translational Modifications (PTMs) are hardly been identified by previous methods because they occur at a lower frequency than canonical amino acids in nature, resulting in lower peptide identification precision. Secondly, there exists various types of noise and missing peaks in the mass spectrum, making the training data (peptide-spectrum matches, PSMs) less reliable. We propose a novel framework, AdaNovo, to calculate the conditional mutual information (CMI) between the spectrum and each amino acid/peptide. And then, we use the CMI to train the model adaptively. Extensive experiments verify that AdaNovo achieve state-of-the-art performance on the 9-species benchmark, where the models are evaluated with never-before-seen peptide labels. Moreover, AdaNovo exhibits significantly superior performance in identifying the PTMs and robustness against data noise. The code for reproducing the results can be found in the supplementary materials.
Tandem mass spectrometry has played a pivotal role in advancing proteomics, enabling the analysis of protein composition in biological samples. Despite the development of various deep learning methods for identifying amino acid sequences (peptides) responsible for observed spectra, challenges persist in \emph{de novo} peptide sequencing. Firstly, prior methods struggle to identify amino acids with post-translational modifications (PTMs) due to their lower frequency in training data compared to canonical amino acids, further resulting in decreased peptide-level identification precision. Secondly, diverse types of noise and missing peaks in mass spectra reduce the reliability of training data (peptide-spectrum matches, PSMs). To address these challenges, we propose AdaNovo, a novel framework that calculates conditional mutual information (CMI) between the spectrum and each amino acid/peptide, using CMI for adaptive model training. Extensive experiments demonstrate AdaNovo's state-of-the-art performance on a 9-species benchmark, where the peptides in the training set are almost completely disjoint from the peptides of the test sets. Moreover, AdaNovo excels in identifying amino acids with PTMs and exhibits robustness against data noise. The supplementary materials contain the official code.
\end{abstract}
\section{Introduction}
\label{intro}
\begin{figure}[t]
    \begin{center}
    \includegraphics[width=0.48\textwidth]{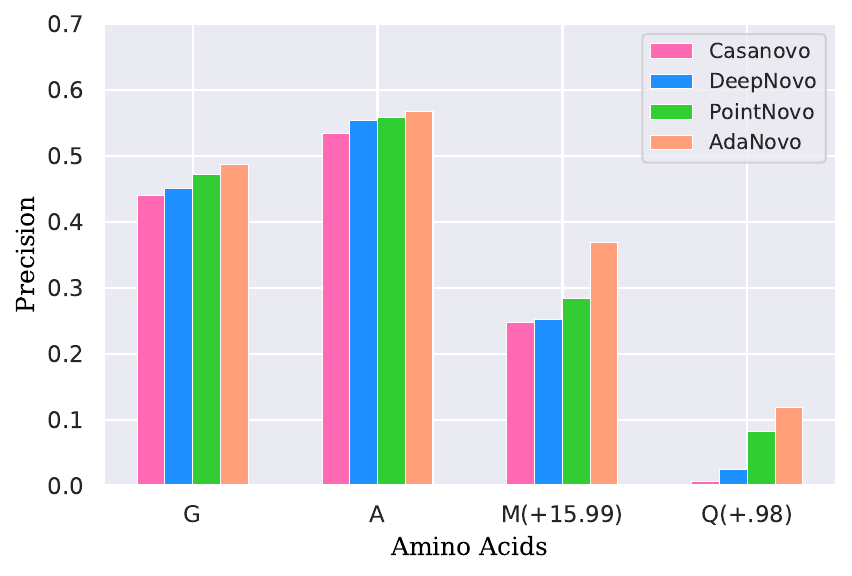}
    \end{center}
    \vspace{-1em}
    \caption{Comparisons of various \emph{de novo} sequencing methods in terms of amino acid-level precision. `G' and `A' denote Glycine and Alanine, respectively. Both of them are canonical amino acids. `M(+15.99)' and `Q(+.98)' represent oxidation of methionine and deamidation of glutamin, both of which are modified amino acids (the amino acids with PTMs). The results are for the human dataset, which is one of 9-species benchmark~\cite{tran2017novo}.}
   \label{ptm}
\end{figure}

\begin{figure*}[t]
    \begin{center}
    \includegraphics[width=0.818\textwidth]{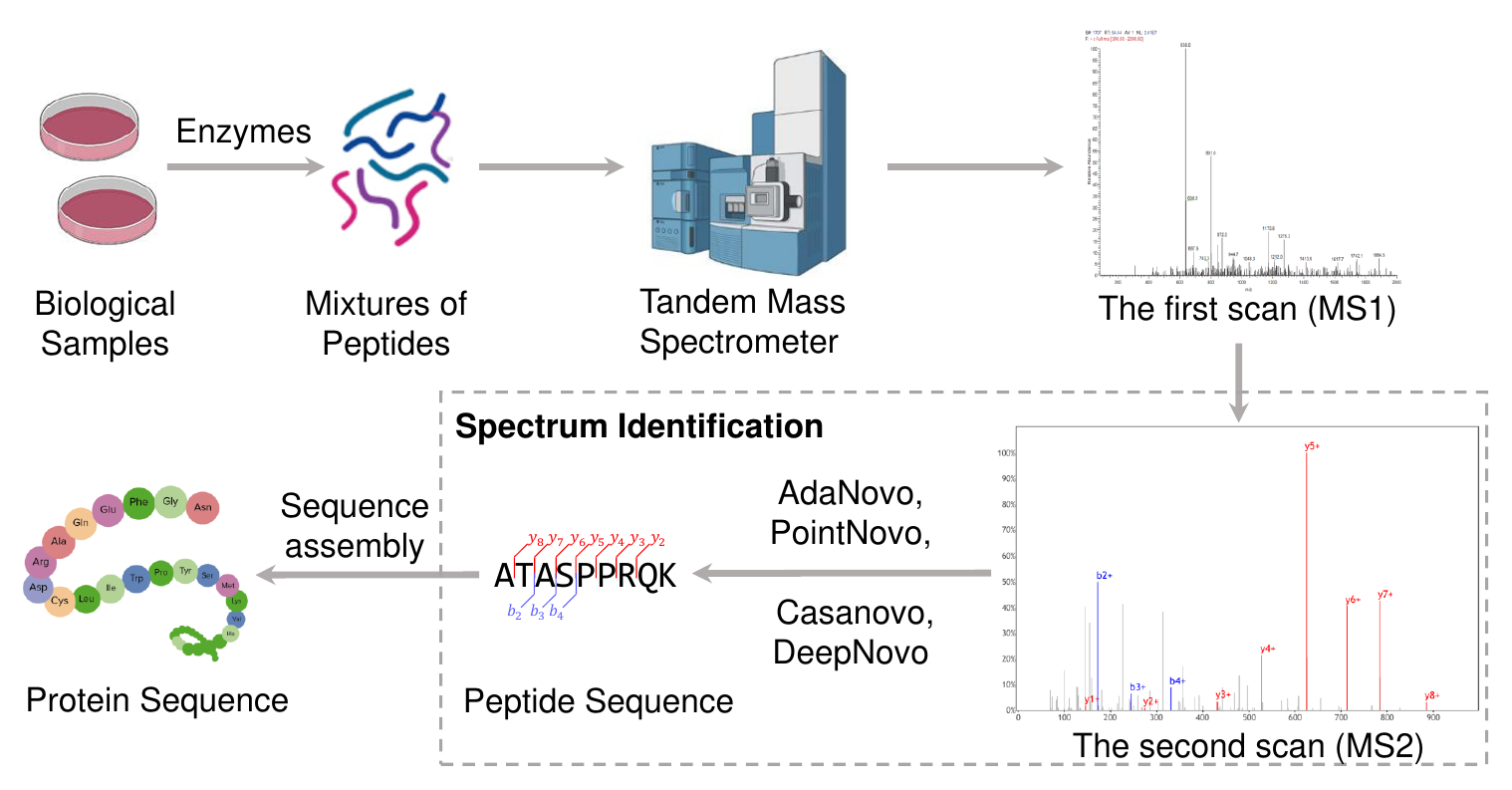}
    \end{center}
    \vspace{-1em}
    \caption{The identification workflow of shotgun proteomics ~\cite{wolters2001automated}. The spectrum identification task this work study is to produce the peptide sequence (e.g., ATASPPRQK) for the observed spectrum. In the spectrum, peaks representing b- and y-ions of the associated peptide are highlighted in color, while grey peaks indicate unexpected fragmentation events or noise. The spectrum annotation are created using ProteomeXchange~\cite{vizcaino2014proteomexchange}.}
   \label{pipeline}
\end{figure*}
Tandem mass spectrometry is a high-throughput tool to identify and quantify proteins in biological samples. However, the precise determination of protein content from observed mass spectra at scale remains a formidable challenge. Central to this challenge is the spectrum identification problem, wherein we are presented with an observed mass spectrum and the corresponding precursor information (mass and charge of the peptide), and our task is to predict the peptide (amino acid sequence) responsible for generating the spectrum. Currently, spectrum identification is most commonly solved using database search, where the observed spectra from the mass spectrometer are compared to theoretical spectra generated by database of known protein sequences. Software algorithms match experimental spectra to theoretical spectra from the database, report the best-scoring peptide-spectrum match (PSM) per spectrum.

However, database search relies on a pre-defined database, preventing the identification of unexpected peptide sequences, such as those originated from genetic variation. Additionally, a database cannot be leveraged for the analysis of some types of immunopeptidomics data~\cite{vanduijn2017immune}, in antibody sequencing~\cite{tran2017novo} or in vaccine development~\cite{mayer2021immunopeptidomics}. Also, the task of constructing a precise database for metaproteomic analyses, including those related to the human microbiome or environmental samples, is deemed impossible~\cite{muth2013searching}. All of these limitations necessitate \emph{de novo} peptide sequencing from the observed mass spectra without using prior knowledge in the form of a peptide sequence database.

Since the early 1990s, \emph{de novo} methods based on the graph theory~\cite{bartels1990fast,frank2009predicting}, Hidden Markov Model~\cite{fischer2005novohmm}, or dynamic programming~\cite{danvcik1999novo,ma2003peaks,frank2005pepnovo} were developed to score peptide sequences against observed spectra. With the rise of deep learning, some researchers train the deep neural networks using PSMs~\cite{tran2017novo,qiao2021computationally,yilmaz2022novo}, where they regard the spectra and matching peptides as the inputs and labels, respectively. And then, the trained models are expected to identify never-before-seen peptides. Although these methods have achieved notable progress, as shown in Figure~\ref{ptm}, we observe that they struggle to identify the amino acids with PTMs, further leading to low amino acid-level and peptide-level precision. However, the identification of amino acids with PTMs holds significant biological importance because PTMs plays a pivotal role in elucidating protein function and studying disease mechanisms~\cite{deribe2010post}. 

On the other hand, some of the expected peaks in mass spectra may be missing due to instrument malfunction or multiple cleavage events occurring on the same peptide, and some additional peaks may undesirably appear in the spectrum, created by instrument noise or non-peptide molecules in the biological samples. All of these make the spectra and peptides labels for training being poorly matched. 

To address above issues, we propose a new framework, AdaNovo, to calculate the conditional mutual information (CMI) between the spectrum and each amino acid in the matching peptide. This can measures the importance of different
target amino acids by their dependence on the source spectrum. Based on the amino acid-level CMI, we obtain the PSM-level CMI between the spectrum and the entire peptide to measure the matching level of each spectrum-peptide pair in the training PSM data. Subsequently, we design an adaptive training approach based on both the amino acid- and PSM-level CMI, which adaptively re-weights the training losses of the corresponding amino acids. 

We conduct the training and evaluation of our model on the widely-used 9-species datasets and observe that AdaNovo outperforms state-of-the-art methods in predicting never-before-seen peptide sequences and demonstrate significantly higher precision in identifying the amino acids with PTMs.
% This is achieved through a robust cross-validation framework, specifically designed for testing on spectra featuring peptide labels that have never been encountered before, involving cross-species prediction. Our experimental findings highlight that Casanovo outperforms state-of-the-art methods in predicting peptide sequences, .  Notably, Casanovo achieves this with inference times that are either equal to or shorter than those of existing methods.

% We train and evaluate our model on a  multi-species bench

% we propose AdaNovo model in this paper which based on conditional mutual information, effectively alleviates both of these challenges in predicting PTMs and the noise present in mass spectrometry data. 

% At the peptide level precision, AdaNovo Significantly exceeding both previous methods based on machine learning across all species, especially in PTM related metrics
% state-of-the-art methods only achieve 36\% \- 57\% peptide-level precision on the widely-used 9-species datasets. 

% \begin{itemize}
% \item 

% \item 

% \item
% \end{itemize}
\section{Background}
Proteomics research focuses on large-scale studies to characterize the proteome, the entire set of proteins, in a living organism. Tandem mass spectrometry (MS), as the mainstream high-throughput technique to identify protein sequences, plays an essential role in proteomics research. As shown in Figure~\ref{pipeline}, in a standard identification workflow of shotgun proteomics ~\cite{wolters2001automated}, proteins undergo initial digestion by enzymes, yielding a mixture of peptides. A tandem mass spectrometer measures mass-to-charge (\emph{m/z}) ratios of each peptides in a two-scan process. During the first scan (MS1), the mass-to-charge (\emph{m/z}) ratios of intact peptides, also known as precursors, are measured. Following this, peptides undergo fragmentation, and the resulting fragments are analyzed in a subsequent scan. In the second scan (MS2), peptides are fragmented at random locations along the peptide backbone, generating peaks corresponding to prefixes (\emph{b-ions}) and suffixes (\emph{y-ions}) of the peptide, each associated with a specific charge state. Consequently, the MS2 spectrum comprises a collection of peaks. Each peak is characterized by an \emph{m/z} value and an associated intensity. The intensity, though unitless, is directly proportional to the number of ions contributing to the observed peak. The \emph{m/z} value is measured with remarkable precision, while the intensity is measured with comparatively lower precision. The core of the above pipeline is the \textbf{spectrum identification} problem, where we aim to predict the peptide sequence responsible for generating the observed MS2 spectrum and the corresponding precursor information (mass and
charge of the peptide) 

\section{Related Work}
Early \emph{de novo} sequencing methods used dynamic programming to score peptide sequences against each observed spectrum. PEAKS \cite{ma2003peaks} uses a sophisticated dynamic programming algorithm to compute the best sequences whose fragment ions can best interpret the peaks in the MS2 spectrum. Graph-based algorithms, such as Sherenga \cite{danvcik1999novo} and pNovo \cite{taylor2001implementation}, first translated the spectrum into a “spectrum graph” where nodes in the graph correspond to peaks in the spectrum and two nodes are connected by an edge if the mass difference between the two corresponding peaks is equal to the mass of an amino acid. The \emph{de novo} peptide sequencing problem is thus cast as finding the path in the resulting graph.

Recently, machine learning \cite{fischer2005novohmm,frank2005pepnovo} have been introduced into \emph{de novo} peptide sequencing and significantly improved the accuracy. The PepNovo \cite{frank2005pepnovo} algorithm present a novel scoring method, which uses a probabilistic network whose structure reflects the chemical and physical rules that govern the peptide fragmentation. The Novor algorithm \cite{ma2015novor} achieved improved performance by using large decision trees as score function in a dynamic programming algorithm.

The first deep neural network method for \emph{de novo} peptide sequencing, DeepNovo \cite{tran2017novo}, treats the \emph{de novo} sequencing task as an image caption task and combines CNN with LSTM to predict the sequence. SMSNet \cite{karunratanakul2019uncovering} is a hybrid approach which leverages a multi-step Sequence-Mask-Search strategy and adopts the encoder-decoder architecture, basically formulating peptide sequencing as a spectra-to-peptide language translation problem. PointNovo \cite{qiao2021computationally} adopts an order invariant network structure for peptide sequencing, which focuses specifically on high-resolution mass spectrometry data. Similar to SMSNet \cite{karunratanakul2019uncovering}, Casanovo \cite{yilmaz2022novo} frames the problem as a language translation problem and employs a transformer framework that has been widely used to process and predict sequences.
 
Although \emph{de novo} methods have achieved notable progress, we observe that they have difficulty in identifying the amino acids with PTMs because these amino acids occur much less frequently in datasets compared to other common amino acids, making it challenging for the model to learn. Additionally, mass spectrometry data contains a significant amount of noise typically originated from the electronic fluctuations in the instruments and other molecules in the biological samples. In other words, there are plenty of noise peaks mixing together with the real ions. All of these make the peptides labels being less reliable. These issues limit the predictive accuracy and widespread use of \emph{de novo} methods. The AdaNovo model proposed in this paper effectively alleviates both of them.
%Initially, .

%With the prosperity of deep learning, researchers try to apply deep neural networks in \emph{De Novo} peptide sequencing. The first work is DeepNovo~\cite{tran2017novo}, which adopts convolutional neural network and long short term memory (LSTM) network to predict the amino acid sequence given the spectrum and precursor. Subsequently, SMSNet algorithm~\cite{karunratanakul2019uncovering} proposes a post-processing step to replace the low-confidence amino acids with peptide database.
%Concurrently, pNovo 3~\cite{yang2019pnovo} uses dynamic programming to generate a set of candidate peptides for a given spectrum and employs pDeep~\cite{} .

%PintNovo~\cite{qiao2021computationally} employs an order-invariant neural network architecture~\cite{qi2017pointnet} to robustly handle any resolution levels of mass spectra while keeping the computational complexity unchanged. More recently, CasaNovo~\cite{yilmaz2022novo} uses a transformer encoder-decoder framework to encode a sequence of observed peaks (a mass spectrum) and decode a sequence of amino acids (a peptide). Despite the remarkable progress, we observe that previous methods face difficulties in identifying PTMs. Also, the PSMs for training may be unreliable because of the instrument noise and the interference from other molecules. Our AdaNovo can mitigate these issues in a unified framework.
\section{Methods}
\subsection{Task Formulation}
Formally, we denote mass spectrum peaks as $\mathbf{x}=\left\{\left(m_i, I_i\right)\right\}_{i=1}^M$, where each peak $\left(m_i, I_i\right)$ forms a 2-tuple representing the $\emph{m/z}$ and intensity value, and $M$ is the number of peaks that can be varied across different mass spectra. Also, we denote the precursor as $\mathbf{z}=\left\{\left(m_{prec}, c_{prec}\right)\right\}$, consisting of the total mass $m_{prec} \in \mathbb{R}$ and charge state $c_{prec} \in \left\{ 1, 2, \dots, 10\right\}$ of the spectrum. Additionally, we represent the peptide sequence as $\mathbf{y}=\left\{\left(y_1, y_2, \dots, y_{N}\right)\right\}$, where $N$ is the peptide length and can be varied across different peptides. $\mathbf{y}_{< j}$ means the previous amino acids sequence appearing before the index $j$ in the peptide $\mathbf{y}$. The \emph{de novo} peptide sequencing models are designed to predict the probability of each amino acid:
\begin{equation}
P(\mathbf{y} \mid \mathbf{x},\mathbf{z} ; \theta)=\prod_{j=1}^{N} p\left(y_j \mid \mathbf{y}_{<j}, \mathbf{x},\mathbf{z} ; \theta\right),
\end{equation}
where $j$ is the index of each amino acid position in the peptide sequence and $\theta$ is the model parameter. In general, previous models~\cite{tran2017novo,yilmaz2022novo,qiao2021computationally} are optimized using the cross-entropy (CE) loss during training:
\begin{equation}
\mathcal{L}_{\mathrm{CE}}(\theta)=-\sum_{j=1}^N \log p\left(y_j \mid \mathbf{y}_{<j}, \mathbf{x},\mathbf{z} ; \theta\right).
\end{equation}
During inference, the \emph{de novo} sequencing models typically predict the probabilities of target amino acids in an autoregressive manner and generate hypotheses using heuristic search algorithms like beam search~\cite{carnegie1977speech}.
% In the \emph{De Novo} peptide sequencing task, we expect to predict \mathbf{y} given the \mathbf{x} and \mathbf{z} from the mass spectrometers. 

\subsection{Model Architectures}
\begin{figure}[t]
    \begin{center}
    \includegraphics[width=0.496\textwidth]{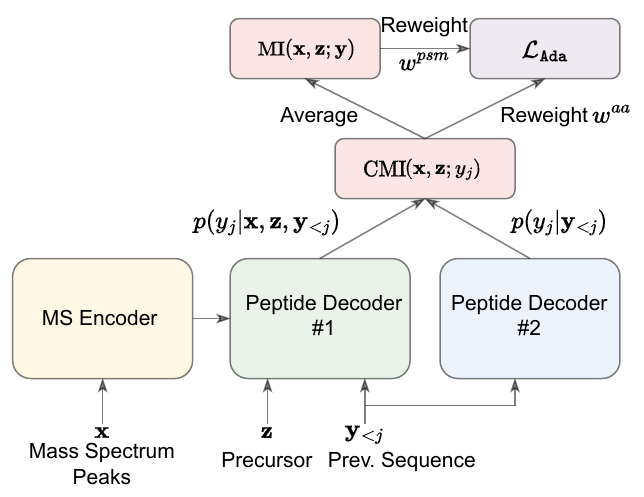}
    \end{center}
    \caption{Schematic diagram of AdaNovo framework.}
    \label{fig_1}
    \vspace{-1em}
\end{figure}

As shown in Figure~\ref{fig_1}, AdaNovo consists of a mass spectrum encoder (\textsf{MS Encoder}) and two peptide decoders (\textsf{Peptide Decoder \#1} and \textsf{Peptide Decoder \#2}). All of these models are built on the Transformer~\cite{vaswani2017attention}.

In order to feed the MS peaks to \textsf{MS Encoder}, we regard each mass spectrum peak $\left(m_i, I_i\right)$ as a `word' in natural language processing and obtain its embedding by individually embed its \emph{m/z} value and intensity value before combining them through summation.  Specifically, we employ a fixed, sinusoidal embedding~\cite{vaswani2017attention} to project $\emph{m/z}$ value $m_i$ to a $d$ dimensional vector $f_i$,
\begin{equation}
f_i=\left\{\begin{array}{ll}
\sin \left(m_i /\left(\frac{\lambda_{\max }}{\lambda_{\min }}\left(\frac{\lambda_{\min }}{2 \pi}\right)^{2 i / d}\right)\right), & \text { for } i \leq d / 2 \\
\cos \left(m_i /\left(\frac{\lambda_{\max }}{\lambda_{\min }}\left(\frac{\lambda_{\min }}{2 \pi}\right)^{2 i / d}\right)\right), & \text { for } i>d / 2
\end{array}\right.
\end{equation}
where $\lambda_{\max}$ = 10,000 and $\lambda_{\min}$ = 0.001. The input embeddings furnish a detailed portrayal of high-precision \emph{m/z} information. Analogous to the consideration of relative positions in the initial transformer model~\cite{vaswani2017attention}, these embeddings potentially facilitate the model's attention to \emph{m/z} variations between peaks. Such attention to detail is crucial for the accurate identification of amino acids within the peptide sequence. The intensity, measured with less precision compared to the \emph{m/z} value, undergoes embedding by projection into d dimensions through a linear layer. Subsequently, the \emph{m/z} and intensity embeddings are amalgamated through summation, resulting in the generation of the input peak embedding. Kindly note that the mass spectrum peaks are permutation invariant, i.e., the order in which the peaks appear in the spectrum does not affect the identification results. Therefore, it is unnecessary to account for an extra positional embedding~\cite{ke2021rethinking} like natural language processing when feed the peaks into transformer. 

Similarly, for the precursor $\mathbf{z}=\left\{\left(m_{prec}, c_{prec}\right)\right\}$ to be fed into the \textsf{Peptide Decoder \#1}, we employ the same sinusoidal embedding for $m_{prec}$ as the \emph{m/z} above and an embedding layer to embed $c_{prec}$. Finally, we obtain the input precursor embedding by summarizing the above 2 embeddings. As for the peptide sequence, the amino acid vocabulary encompasses the 20 canonical amino acids, along with post-translationally modified versions of three among them (oxidation of methionine and deamidation of asparagine or glutamine). Additionally, a special \texttt{stop} token signals the end of decoding, resulting in a total of 24 tokens. \textsf{Peptide Decoder \#1} and \textsf{Peptide Decoder \#2} undergo autoregressive training, wherein they receive the preceding amino acid sequence $\mathbf{y}_{<j}$ prior to amino acid $j$ during the prediction process for the identity of amino acid $j$. However, different from \textsf{Peptide Decoder \#1}, \textsf{Peptide Decoder \#2} exclusively utilizes $\mathbf{y}_{<j}$ as input because we want to calculate the conditional probability $p(y_{j} \mid \mathbf{y}_{<j})$, which is the prerequisite for calculating the conditional mutual information between the mass spectrum ($\mathbf{x}$ and $\mathbf{z}$) and amino acid $y_{j}$.
% exclude the information from the MS ($\mathbf{x}$ and $\mathbf{z}$).
% It is noteworthy that the Peptider \#1 takes .
% Additionally, the \textsf{Peptide Decoder \#1} also .
% To obtain the peak embedding, we individually embed the \emph{m/z value and intensity value before combining them through summation. Specifically, for the .
% and 
% feed the mass spectrum peaks set to the MS Encoder . Kindly note that .
% Next, we describe how to feed above data into the transformer models. 
% The $\emph{m/z}$ value and intensity undergo separate embeddings before being combined to create the input peak embedding. For the projection of each $\emph{m/z}$ value to a $d$ dimensional vector, we employ a fixed, sinusoidal embedding method (Vaswani et al., 2017). This method involves generating the $\emph{m/z}$ embedding $f$ from an equal number of sine and cosine waveforms covering wavelengths from 0.001 to 10,000 $\mathrm{~\emph{m/z}}$. Each feature in the embedding $f_i$ is then crafted as a distinctive element in this embedding space,
% \begin{equation}
% f_{i, j}=\left\{\begin{array}{l}
% \sin \left(\frac{m_i}{\frac{\lambda_{\min }}{2 \pi}\left(\frac{\lambda_{\max }}{\lambda_{\min }}\right)^{2 j / d}}\right), j=1, \cdots, \frac{d}{2} ; \\
% \cos \left(\frac{m_i}{\frac{\lambda_{\min }}{2 \pi}\left(\frac{\lambda_{\max }}{\lambda_{\min }}\right)^{2 j / d}}\right), j=\frac{d}{2}+1, \cdots, d ;
% \end{array}, i=1, \cdots, N\right.
% \end{equation}
\subsection{Training Strategies}
The training strategies consist of amino acid-level (\S~\ref{aa-level}) and PSM-level adaptive training (\S~\ref{psm-level}), which we elaborate on below.
\subsubsection{Amino acid-level Adaptive Training.}
\label{aa-level}
As mentioned above, previous \emph{de novo} sequencing models struggle to identify amino acids with PTMs because they occur much less frequently in datasets compared to other canonical amino acids, making it challenging for the model to learn. Therefore, we expect to emphasize the amino acids with PTMs to improve the models' ability in identifying them. This resembles the up-sampling methods in long-tailed classification where researchers emphasize samples from the tail class during training~\cite{zhang2023deep,ren2018learning}. We also compare with these alternative methods in Section~\ref{alter_w}. On the other hand, when predict the amino acid with PTMs $y_j$, we should rely more on mass spectrometry data (\textrm{x} and \textrm{z}) and less on the historical predictions of previous amino acids $y_{<j}$ because the mass shifts resulting from PTMs are only manifested in the mass spectrometry data. This motivates us to measure the mutual information (MI) between each target amino acid and mass spectrum conditioned on previous amino acids, i.e., conditional mutual information (CMI)~\cite{wyner1978definition} between each target amino acid and mass spectrum,
\[\begin{aligned}
 \operatorname{CMI} (\mathbf{x}, \mathbf{z}; y_j)&= \operatorname{MI} \left(\mathbf{x}, \mathbf{z}; y_j\mid \mathbf{y}_{<j} \right) \\& =\log \left(\frac{p\left(y_j, \mathbf{x}, \mathbf{z}\mid \mathbf{y}_{<j}\right)}{p\left(y_j \mid \mathbf{y}_{<j}\right) \cdot p\left(\mathbf{x}, \mathbf{z}\mid \mathbf{y}_{<j}\right)}\right).
\end{aligned}\]
However, it is computationally impractical to calculate the $\operatorname{CMI}$ with above definition. To address this, we proceed to enhance its computational tractability by decomposing the conditional joint distribution,
\[\begin{aligned}
 \operatorname{CMI} (\mathbf{x}, \mathbf{z}; y_j)&= \operatorname{MI} \left(\mathbf{x}, \mathbf{z}; y_j\mid \mathbf{y}_{<j} \right) \\& =\log \left(\frac{p\left(y_j, \mathbf{x}, \mathbf{z}\mid \mathbf{y}_{<j}\right)}{p\left(y_j \mid \mathbf{y}_{<j}\right) \cdot p\left(\mathbf{x}, \mathbf{z}\mid \mathbf{y}_{<j}\right)}\right). \\
& =\log \left(\frac{p\left(y_j \mid \mathbf{x}, \mathbf{z},\mathbf{y}_{<j}\right) \cdot p\left(\mathbf{x},\mathbf{z} \mid \mathbf{y}_{<j}\right)}{p\left(y_j \mid \mathbf{y}_{<j}\right) \cdot p\left(\mathbf{x},\mathbf{z} \mid \mathbf{y}_{<j}\right)}\right) \\
& =\log \left(\frac{p\left(y_j \mid \mathbf{x},\mathbf{z}, \mathbf{y}_{<j}\right)}{p\left(y_j \mid \mathbf{y}_{<j}\right)}\right).\\
\end{aligned}\]
In this way, the $\operatorname{CMI}(\mathbf{x}, \mathbf{z}; y_j)$ can be obtained with $p\left(y_j \mid \mathbf{x},\mathbf{z}, \mathbf{y}_{<j}\right)$ and $p\left(y_j \mid \mathbf{y}_{<j}\right)$, which are the output of the \textsf{Peptide Decoder \#1} and \textsf{Peptide Decoder \#2}, respectively. 
Moreover, to reduce the variances and stabilize the distribution of the amino acid-level CMI in each peptide, we normalize the CMI values in the peptide and then scale the normalized values to obtain the amino acid-level training weight $w_j^{aa}$ for $y_j$,
\begin{equation}
w_j^{aa}=\max \left\{0, s_{1} \cdot \frac{\operatorname{CMI}\left(\mathbf{x}, \mathbf{z}; y_j\right)-\mu^{aa}}{\sigma^{aa}} +1\right\},
\end{equation}
where $\mu^{aa}$ and $\sigma^{aa}$ are the mean values and the standard deviations of all the CMI values in each peptide, and $s_{1}$ is a hyperparameter that controls the effect of amino acid-level adaptive training.
\subsubsection{PSM-level Adaptive Training.}
\label{psm-level}
As we introduced before, the training PSMs samples are of different matching levels because of the signal noise and missing peaks. To alleviate the negative effect of poorly matched mass spectrometry and peptide pairs and encourage the well-matched pairs, we adopt the mutual information between them as a measure of matching levels. Formally,
% In this section, we first introduce the definition of conditional mutual information and the model architectures of MS encoder and peptide decoder. Then, we illustrate how to adjust the weights for the training losses based on the amino acid-level (Sec.~\ref{aa}) and PSM-level (Sec.~\ref{psm}) conditional mutual information. Figure~\ref{fig_1} shows the overall training process of our approach.
% \subsection{Amino acid-level CMI}
% \label{aa}
% \begin{equation}
% \begin{aligned}
%  I_{j}&= I\left(\mathbf{x}, \mathbf{z}; y_j\mid \mathbf{y}_{<j} \right) \\& =\log \left(\frac{p\left(y_j, \mathbf{x}, \mathbf{z}\mid \mathbf{y}_{<j}\right)}{p\left(y_j \mid \mathbf{y}_{<j}\right) \cdot p\left(\mathbf{x}, \mathbf{z}\mid \mathbf{y}_{<j}\right)}\right) \\
% & =\log \left(\frac{p\left(y_j \mid \mathbf{x}, \mathbf{z},\mathbf{y}_{<j}\right) \cdot p\left(\mathbf{x},\mathbf{z} \mid \mathbf{y}_{<j}\right)}{p\left(y_j \mid \mathbf{y}_{<j}\right) \cdot p\left(\mathbf{x},\mathbf{z} \mid \mathbf{y}_{<j}\right)}\right) \\
% & =\log \left(\frac{p\left(y_j \mid \mathbf{x},\mathbf{z}, \mathbf{y}_{<j}\right)}{p\left(y_j \mid \mathbf{y}_{<j}\right)}\right) ,\\
% % & =\log \left(\frac{p_{\mathrm{NMT}}\left(y_j\right)}{p_{\mathrm{LM}}\left(y_j\right)}\right)
% w_j^{aa}= \operatorname{SoftMax}(I_{j})
% \end{aligned}
% \end{equation}
% \tilde{w}_{i, t}=\max \left(u_{i, t}, 0\right),
% w_{i, t}=\frac{\tilde{w}_{i, t}}{\left(\sum_j \tilde{w}_{j, t}\right)+\delta\left(\sum_j \tilde{w}_{j, t}\right)}
% \delta(a)=1 \text { if } a=0
% 
% \subsection{PSM-level CMI}
% \label{psm}
\begin{equation}
\begin{aligned}
\operatorname{MI}(\mathbf{x}, \mathbf{z}; \mathbf{y}) & =\frac{1}{|\mathbf{y}|} \log \left(\frac{p(\mathbf{x}, \mathbf{z}, \mathbf{y})}{p(\mathbf{x}, \mathbf{z}) \cdot p(\mathbf{y})}\right) \\
& =\frac{1}{|\mathbf{y}|} \log \left(\frac{p(\mathbf{y} \mid \mathbf{x}, \mathbf{z})}{p(\mathbf{y})}\right) \\
& =\frac{1}{|\mathbf{y}|} \log \left(\frac{\prod_j p\left(y_j \mid \mathbf{x}, \mathbf{z}, \mathbf{y}_{<j}\right)}{\prod_j p\left(y_j \mid \mathbf{y}_{<j}\right)}\right) \\
& =\frac{1}{|\mathbf{y}|} \sum_j \log \left(\frac{p\left(y_j \mid \mathbf{x}, \mathbf{z}, \mathbf{y}_{<j}\right)}{p\left(y_j \mid \mathbf{y}_{<j}\right)}\right) \\
& =\frac{1}{|\mathbf{y}|} \sum_j \operatorname{CMI} (\mathbf{x}, \mathbf{z}; y_j).
\end{aligned}
\end{equation}
In other words, the mutual information can be derived by averaging all the amino acid-level $\operatorname{CMI} (\mathbf{x}, \mathbf{z}; y_j)$ over the peptide. Similarly, we normalize all the $\operatorname{MI}$ values across all the PSMs in each mini-batch and then scale the normalized values to obtain the PSM-level training weight $w^{psm}$,
\begin{equation}
w^{psm}=\max \left\{0, s_{2} \cdot \frac{\operatorname{MI}(\mathbf{x}, \mathbf{z}; \mathbf{y})-\mu^{psm}}{\sigma^{psm}} +1\right\},
\end{equation}
where $\mu^{psm}$ and $\sigma^{psm}$ are the mean values and the standard deviations of the MI values of all the PSMs in each minibatch, and $s_{2}$ is a hyperparameter that controls the effect of PSM-level adaptive training.
\subsubsection{Adaptive Training Loss}
In our adaptive training method, we re-weight each target amino acid $y_j$ with the following loss,
\begin{equation}
\mathcal{L}_{1}(\theta_1)=-\sum_{j=1}^N w_j \log p\left(y_j \mid \mathbf{y}_{<j}, \mathbf{x}, \mathbf{z}; \theta_1\right),
\end{equation}
where $\theta_1$ are the parameters of \textsf{MS Encoder} and \textsf{Peptide Decoder \#1}, and
\begin{equation}
w_j=w_j^{aa} \cdot w^{psm}.
\end{equation}
Additionally, \textsf{Peptide Decoder \#2} is trained with the following loss,
\begin{equation}
\mathcal{L}_{2}(\theta_2)=-\sum_{j=1}^{N} \log p\left(y_j \mid \mathbf{y}_{<j}; \theta_2\right),
\end{equation}
where  $\theta_2$ are the parameters of \textsf{Peptide Decoder \#2}.
The overall training loss is,
\begin{equation}
\mathcal{L}_{\tt{Ada}}(\theta_1, \theta_2)= \mathcal{L}_{1}(\theta_1) + \mathcal{L}_{2}(\theta_2).
\end{equation}

\subsection{Inference}
In the inference phase, we only use \textsf{MS Encoder} and \textsf{Peptide Decoder \#1} to predict the peptide. Specifically, we feed the mass spectrometry data to the encoder \textsf{MS Encoder} and the decoder \textsf{Peptide Decoder \#1} predicts the highest-scoring amino acid for each peptide sequence position. The decoder is then fed its preceding amino acid predictions at each decoding step. The decoding process concludes upon predicting the stop token or reaching the predefined maximum peptide length of $\ell=100$ amino acids. We discuss the computational overhead in Section~\ref{cost}.

\subsection{Precursor \emph{m/z} filtering}
In \emph{de novo} peptide sequencing, it's crucial that the relative difference between the total mass of the predicted peptide ($m_{\text{pred}}$) and the observed precursor mass ($m_{\text{prec}}$) remains below a specified threshold value $\epsilon$ for the predicted sequence to be considered plausible. This requirement is expressed as $\Delta m_{ppm}=\frac{\left|m_{\text{prec}}-m_{\text{pred}}\right| \times 10^6}{m_{\text{prec}}} < \epsilon$. To ensure adherence to this constraint, we not only incorporate precursor information into the model's learning process but also filter out peptide predictions that don't meet this criterion. The threshold value $\epsilon$ is determined based on the precursor mass error tolerance used in the database search to establish ground truth peptide sequences for the test data.

\section{Experiments}
\begin{table*}[ht]
    \centering
    \caption{Empirical comparison of \emph{de novo} sequencing models. The table lists the peptide-level and amino acid-level precision of three competing models on all nine benchmark cross-validation folds. Each fold's test set contains spectra from a single species. \textbf{Kindly note that peptide-level performance measures are the primary quantifier of the model’s practical utility because the goal is to assign a complete peptide sequence to each spectrum.} The best and the second best results are highlighted \textbf{bold} and \underline{underlined}, respectively.}
\setlength{\tabcolsep}{3pt}
    \centering
    \fontsize{7.6pt}{\baselineskip}\selectfont
\begin{tabular}{l|cccc|cccc}
\toprule \multirow[c]{2}{*}{ Species } & \multicolumn{4}{c}{Peptide-level precision}  & \multicolumn{4}{|c}{Amino acid-level precision} \\
 &  
DeepNovo  & 
PointNovo   & Casanovo & AdaNovo  & 
DeepNovo 
 & 
PointNovo & Casanovo &AdaNovo\\
\midrule
 Mouse & 0.286 & 0.355  & \underline{0.449} & \textbf{0.467}& 0.623 & \underline{0.626}  & 0.612 & \textbf{0.646}\\
 Human & 0.293 & \underline{0.351} & 0.343 &\textbf{0.373} & \underline{0.610}& 0.606&0.585 &\textbf{0.618} \\
 Yeast & 0.462 & 0.534 &\underline{0.568} &\textbf{0.593}  & 0.750 & \underline{0.779} & 0.753& \textbf{0.793} \\
 \emph{M. mazei} & 0.422 & \underline{0.478}  & 0.474& \textbf{0.496} & 0.694 & \underline{0.712}  & 0.686 &\textbf{0.728} \\
 Honeybee & 0.330 & 0.396 &\underline{0.422} & \textbf{0.431}& 0.630 & \underline{0.644}  & 0.640& \textbf{0.650} \\
 Tomato & 0.454 & \underline{0.513}  & 0.463& \textbf{0.530}& 0.731 & \underline{0.733}  & 0.720 &\textbf{0.740}\\
 Rice bean & 0.436 &0.511  & \textbf{0.549}& \underline{0.546} &0.679 &\textbf{0.730} &\underline{0.727} & 0.719\\
 Bacillus & 0.449 & \underline{0.518} & 0.513 & \textbf{0.528}  & \underline{0.742} & \textbf{0.768}  & 0.718 & 0.739\\
 Clam bacteria & 0.253 & 0.298 & \underline{0.347}&\textbf{0.372} & 0.602 & 0.589 & \underline{0.617} &\textbf{0.642} \\
\bottomrule
\end{tabular}
\label{main}
\end{table*}

\subsection{Datasets}
To assess AdaNovo's performance, we employ the nine-species benchmark initially introduced by DeepNovo~\cite{tran2017novo}. This dataset amalgamates approximately 1.5 million mass spectra from nine distinct experiments, each employing the same instrument to scrutinize peptides from diverse species. Each spectrum is associated with a ground-truth peptide sequence, which comes from database search identification with a standard false discovery rate (FDR) set at 1\%. Following the methodology of previous works~\cite{tran2017novo,qiao2021computationally}, we adopt a leave-one-out cross-validation framework. This entails training a model on eight species and testing it on the species held out for each of the nine species. The training set is split 90/10 for training and validation. This framework facilitates the testing of the model on peptide samples that have never been encountered before. Cross-species testing is of paramount importance for \emph{de novo} sequencing models since practical applications often demand these models to excel in handling mass spectra featuring peptide sequences that have never been observed before.
\subsection{Evaluation Metrics}
In our assessment of model predictions, we employ precision calculated at both the amino acid and peptide levels, following methodologies presented by previous works~\cite{ma2003peaks,tran2017novo}. These precision metrics serve as performance measures, gauging the quality of a given model's predictions based on coverage over the test set. For each spectrum, we compare the predicted sequence to the ground truth peptide obtained from the database search.

Consistent with DeepNovo~\cite{tran2017novo}, our approach to amino acid-level measures begins by calculating the number $N_{\text {match}}^a$ of matched amino acid predictions. These are defined as predicted amino acids that (1) exhibit a mass difference of $<0.1 \mathrm{Da}$ from the corresponding ground truth amino acid and (2) have either a prefix or suffix with a mass difference of no more than $0.5 \mathrm{Da}$ from the corresponding amino acid sequence in the ground truth peptide. Amino acid-level precision is then defined as $N_{\text {match }}^a / N_{\text {pred }}^a$, where $N_{\text {pred}}^a$ represents the number of predicted amino acids. Similarly, amino acids with PTMs identification precision can be formulated as $N_{\text {match }}^{ptm} / N_{\text {pred }}^{ptm}$, where $N_{\text {match }}^{ptm}$ and $N_{\text {pred}}^{ptm}$ denote the number of matched amino acids with PTMs and predicted amino acids with PTMs, respectively.

For peptide predictions, a predicted peptide is deemed a correct match only if all of its amino acids are matched. In a collection of $N_{\text {orig }}^p$ spectra, if our model provides predictions for a subset of $N_{\text {pred }}^p$ and accurately predicts $N_{\text {match }}^p$ peptides, coverage is defined as $N_{\text {pred }}^p / N_{\text {orig }}^p$. Peptide-level precision is calculated as $N_{\text {match }}^p / N_{\text {pred }}^p$.

To construct a precision-coverage curve, predictions are sorted based on the confidence score provided by the model. Amino acid-level confidence scores are derived by applying a softmax to the transformer decoder's output, serving as a proxy for the probability of each predicted amino acid occurring at a specific position along the peptide sequence. AdaNovo provides amino acid-level confidence scores directly, and we utilize the mean score across all amino acids as a peptide-level confidence score.
\subsection{Baselines}
We compare AdaNovo with previous \emph{de novo} peptide sequencing methods including DeepNovo~\cite{tran2017novo}, Casanovo~\cite{yilmaz2022novo} and PointNovo~\cite{qiao2021computationally}. We reproduce the results of Casanovo with the settings and hypermeters of the original paper and report the published results of DeepNovo and PointNovo as their pre-trained weights are unavailable.
\subsection{Experimental Settings}
The models in our AdaNovo are with 9 layers, embedding size $d$ = 512, and 8 attention heads. We train the models with a batchsize of 32 PSMs and $10^{-5}$ weight decay. The learning rate is linearly increased from zero to $5 \times 10^{-4}$ in 100$k$ warm-up steps, followed by a cosine
shaped decay. We train the models for 30 epochs and pick the model weights from the epoch with the lowest validation loss for testing. The hypermeters $s_1$ and $s_2$ are tuned within the set $\left\{0.05, 0.1, 0.3\right\}$.
\begin{figure*}[ht]
    \subfigure[Human (Peptide-level)]{
    \label{fig4-a}
    \includegraphics[width=0.235\textwidth]{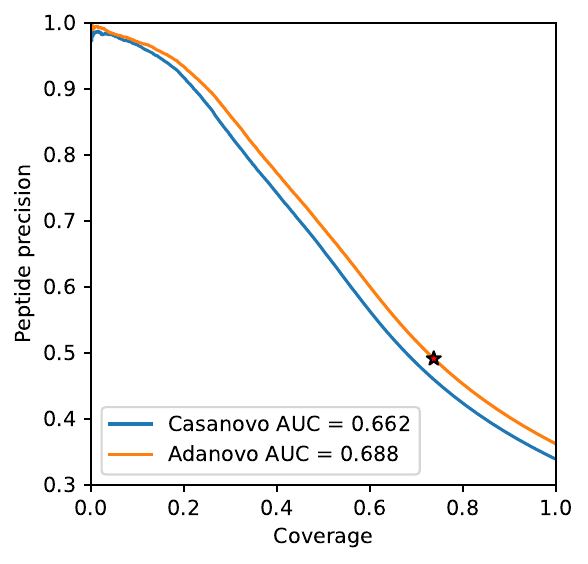}}
    \subfigure[Human (AA-level)]{
    \label{fig4-b}
    \includegraphics[width=0.235\textwidth]{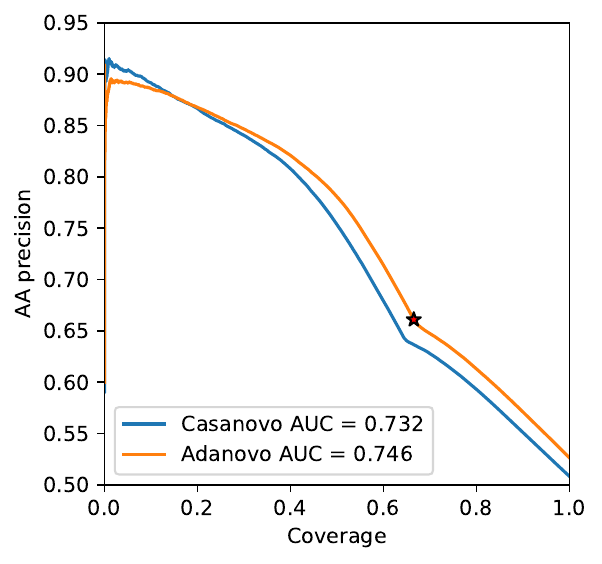}}
    \subfigure[Clam bacteria (Peptide-level)]{
    \label{fig4-c}
    \includegraphics[width=0.235\textwidth]{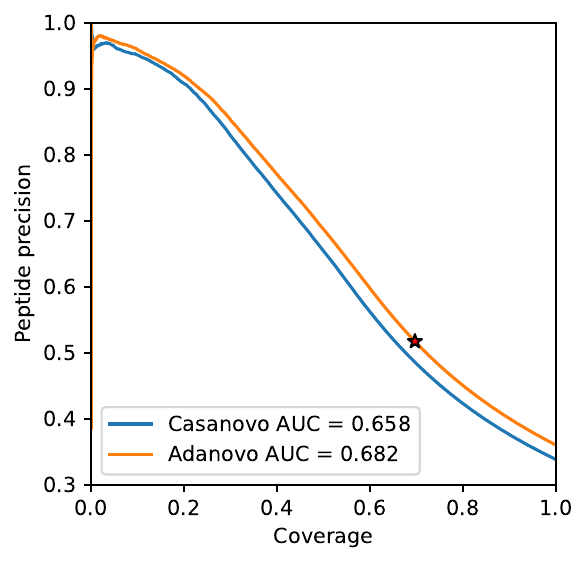}}
    \subfigure[Clam bacteria (AA-level)]{
    \label{fig4-d}
    \includegraphics[width=0.235\textwidth]{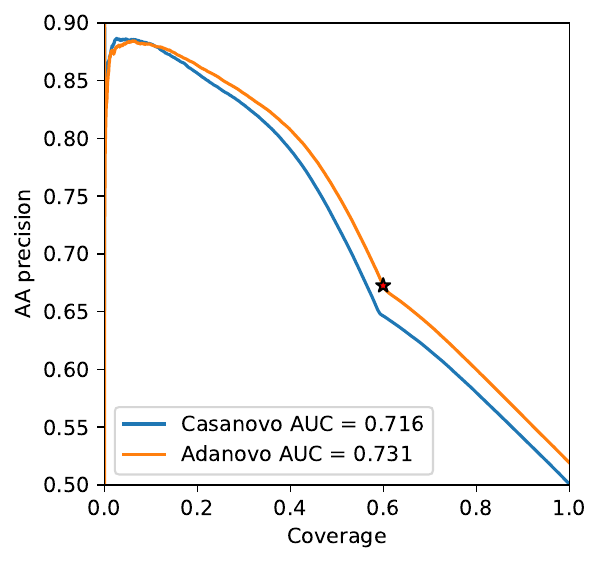}}
    \caption{Precision-coverage curves for AdaNovo and Casanovo (AA-level: Amino acid-level). Peptide curves are generated by arranging predicted peptides based on their confidence scores. In the case of amino acid-level curves, all amino acids within a specific peptide are assigned equal scores. Both at the amino acid and peptide levels, peptides that meet the precursor \emph{m/z} filtering criteria are prioritized over those that do not. Similarly, the ranking is applied to all amino acids within peptides that pass the precursor \emph{m/z} filter compared to those that do not. The transition between unfiltered and filtered entries is denoted by a red star on each curve.}
    \label{fig4}
\end{figure*}
\subsection{Main Results}
\textbf{AdaNovo outperforms state-of-the-art methods.} As can be observed in Table~\ref{main}, AdaNovo outperforms competitive models on most (8 out of 9) species in peptide-level precision compared to DeepNovo, PointNovo and CasaNovo. The peptide-level precision coverage curves (Figure~\ref{fig4}) show that AdaNovo consistently outperforms Casanovo over a range of peptide confidence thresholds. This trend is also reflected by the area under the curve (AUC) metric. At amino acid-level, AdaNovo outperforms baselines on most datasets. As shown in Figure~\ref{fig4}, the point on the AdaNovo curve corresponding to the filter lies above the casanovo precision-coverage curve, and Adanovo’s AUC consistently exceeds Casanovo’s.
\begin{table}[t]
    \centering
    \caption{Empirical comparison of \emph{de novo} sequencing models in terms of identifying amino acids with PTMs. The best and the second best results are highlighted \textbf{bold} and \underline{underlined}, respectively.}
\setlength{\tabcolsep}{3pt}
    \centering
    \fontsize{7.6pt}{\baselineskip}\selectfont
\begin{tabular}{l|cccc}
\toprule \multirow[c]{2}{*}{ Species }  & \multicolumn{4}{|c}{PTMs precision} \\
 &  
DeepNovo  & 
PointNovo   & Casanovo & AdaNovo  \\
\midrule
 Human & 0.369 & \underline{0.415}  & 0.398 &\textbf{0.483}  \\
 Rice bean & 0.644 & \underline{0.653} & 0.646 & \textbf{0.689}  \\
 Clam bacteria & 0.510 & \underline{0.526}  &0.508  & \textbf{0.575}  \\
 Bacillus & 0.483 & \underline{0.524}  &0.470  & \textbf{0.565}  \\
\bottomrule
\end{tabular}
\label{ptms}
\end{table}

\bigskip
\textbf{AdaNovo can accurately identify the amino acids with PTMs.} As demonstrated in Table~\ref{ptms}, we compare AdaNovo with other methods in terms of identifying amino acids with PTMs because AdaNovo is designed to accurately identify the amino acids with PTMs. The results in the table indicate that AdaNovo exceeds other competitors by significant margins in identifying amino acids with PTMs, verifying the effectiveness of the amino acid-level adaptive training strategy.
\subsection{Ablation Study}

\begin{table}[h]
    \centering
    \small
    \caption{Ablations on amino acid-level (AA-level) and peptide-level adaptive training strategies. The results are for the Human test set, which is one of 9-species benchmark~\cite{tran2017novo}.}
    \setlength{\tabcolsep}{3pt}
    \centering
    \fontsize{7.9pt}{\baselineskip}\selectfont
    \begin{tabular}{c|c|c|c}
    \toprule Model & AA. Prec. & Peptide Prec. & PTMs Prec. \\
    \midrule 
    Casanovo & 0.585 & 0.343 & 0.300\\
   AdaNovo (w/o PSM-level MI) & 0.607  & 0.360 & 0.478\\
   AdaNovo (w/o AA-level CMI) & 0.594 & 0.349  & 0.314\\
    AdaNovo & 0.618 &0.373 & 0.483\\
    \bottomrule
    \end{tabular}
    \label{ablation}
\end{table}
\textbf{Ablations on amino acid-level and peptide-level adaptive training strategies.} To investigate the influence of the amino acid-level and peptide-level adaptive training strategies, we remove each of them from AdaNovo. The results shown in Table~\ref{ablation} indicate that both modules are necessary and effective for the AdaNovo model. Moreover, when we remove the AA-level training strategy in AdaNovo, the precision of the amino acids with PTMs identification drops significantly because the amino acid-level training strategy is designed for identifying amino acids with PTMs. Additionally, the PSM-level training strategy is designed for robustness against data noise, which we verify via the following experiments.

\begin{table}[h]
    \centering
    \small
    \caption{Models' Performance on mass spectrum dataset with synthetic noise. The results are for the Clam bacteria test set, which is one of 9-species benchmark~\cite{tran2017novo}.}
    \setlength{\tabcolsep}{3pt}
    \centering
    \fontsize{8.6pt}{\baselineskip}\selectfont
    \begin{tabular}{c|c|c}
    \toprule Model & AA. Prec. & Peptide Prec. \\
    \midrule CasaNovo & 0.582  & 0.297  \\
   AdaNovo (w/o PSM-level MI) & 0.586 & 0.311 \\
   AdaNovo (w/o AA-level CMI) & 0.614 & 0.335  \\
    AdaNovo & 0.621& 0.342 \\
    \bottomrule
    \end{tabular}
    \label{noise}
\end{table}
\textbf{Performance on mass spectra with synthetic noise.}
To verify the effectiveness of the PSM-level adaptive training strategy, we randomly choose 20\% spectrum in the training datasets, and add synthetic noise peaks or remove original peaks with higher intensity values. We report the results in Table~\ref{noise}, from which we can observe that the performance would degrade sharply when we remove the PSM-level training strategy. This indicates that PSM-level adaptive training strategy can enhance models' robustness against data noise in mass spectrum.

\subsection{Comparisons with Alternative Methods for identifying amino acids with PTMs}
\label{alter_w}
\begin{table}[h]
    \centering
    \small
    \caption{Comparisons with alternative methods in terms of identifying amino acids with PTMs. All results are for the yeast test set, which is one of 9-species benchmark~\cite{tran2017novo}.}
    \begin{tabular}{l|c|c}
    \toprule Model & AA. Prec. & Peptide Prec. \\
    \midrule Casanovo & 0.753 & 0.568 \\
   \qquad + Re-weight &0.762 & 0.576 \\
   \qquad + Focal loss & 0.745 & 0.543\\
    AdaNovo (w/o PSM-level MI) & 0.784 &0.582 \\
    AdaNovo & 0.793 &0.593 \\
    \bottomrule
    \end{tabular}
    \label{alter}
\end{table}
In this section, we show the performance of AdaNovo only with amino acid-level loss (denoted as ` AdaNovo w/o PSM-level MI') and compare to some alternative methods in terms of identifying amino acids with PTMs. The first alternative is to re-weight each amino acid $y_{j}$ with the following function,
\begin{equation}
    w_{j} = \frac{N_{total}}{N_{y_j}},
\end{equation}
where $N_{total}$ and $N_{y_j}$ represent the total number of amino acids and the number of amino acids in the $y_j$ category in the dataset, respectively. The second alternative is the focal loss~\cite{lin2017focal}, we replace the cross entropy loss of Casanovo~\cite{yilmaz2022novo} with the focal loss,\[\mathcal{L} = -(1-\alpha p(y_j\mid \mathbf{x}, \mathbf{z}, \mathbf{y}_{<j}))^\gamma \log p(y_j\mid \mathbf{x}, \mathbf{z}, \mathbf{y}_{<j}),\]where $\alpha$ and $\gamma$ are hyperparameters to adjust the loss weight. The results shown in Table~\ref{alter} indicate that both AdaNovo and the first alternative can help improve Casanovo's ability. Additionally, AdaNovo outperforms the alternatives by a notable margin probably because the training and testing datasets are derived from different species, there exists a significant difference in the distribution of PTMs quantities. Also, AdaNovo is inspired by the domain knowledge that the mass shift of PTMs only manifests in the mass spectra, thus shows superiority over the re-weighting methods in long-tailed classification. 

\subsection{Sensitivity Analysis}
The effects of two hyperparameters s1 and s2, which determines the influence of amino acid-level and PSM-level training strategy can be seen in Appendix \ref{sec:appendix_para}.

\subsection{Costs of Computing and Storage}
The comparison between AdaNovo and Casanovo regarding model parameters and runtime can be found in Appendix \ref{sec:appendix_cost}.

\label{cost}

\section{Conclusion and Future Work}
In this paper, we discern challenges in existing methods related to identification of the amino acids with PTMs, exacerbated by spectrum data noise stemming from instrument malfunctions and contaminants. These challenges contribute to reduced precision in identification. To address these issues, we introduce a novel approach involving the calculation of conditional mutual information between the spectrum and each amino acid, followed by a re-weighting of each amino acid. Extensive experiments on widely-utilized 9-species datasets affirm that AdaNovo surpasses previous \emph{de novo} sequencing methods, showcasing superior performance in both amino acid- and peptide-level precision. Notably, AdaNovo exhibits a distinct advantage in identifying amino acids with PTMs. In the future, we plan to train the AdaNovo model on more extensive PSM data, positioning it as a foundational model for mass spectrum-based proteomics.
\section{Impact Statements}
This paper presents work whose goal is to advance the field of machine learning for protein sequencing. There are many potential societal consequences of our work, none of which we feel must be specifically highlighted here.

% \nocite{langley00}
\bibliography{main}
\bibliographystyle{icml2024}
\newpage
\appendix
\onecolumn
\section{Costs of Computing and Storage}
In this part, we compare AdaNovo with Casanovo in
terms of the number of model parameters, training time and
inference time. The results shown in Table ~\ref{cost_table}. Compared to
casanovo, AdaNovo introduced \textsf{Peptide Decoder \#2}, resulting in a 40.04\% increase in parameter count (from 47.35M
to 66.31M). Similarly, under the same hardware settings (1
A100-SXM4-80GB and 32 CPU), training time increased
by 7.3\% (from 63.27M to 67.92M). However, the inference of AdaNovo is more efficient than CasaNovo.
\begin{table}[h]
    \centering
    \small
    \caption{Comparisons with competitive methods in terms of computational overhead. The training and inference time are evaluated on Honeybee dataset, which is one of 9-species benchmark~\cite{tran2017novo}.}
    \setlength{\tabcolsep}{3pt}
    \centering
    \fontsize{7.6pt}{\baselineskip}\selectfont
    \begin{tabular}{l|c|c|c}
    \toprule Model & \textbf{\#Params (M)} & \textbf{Training time (h)} & \textbf{Inference time (h)}\\
    \midrule 
    % DeepNovo &  &  &\\
     Casanovo & 47.35 & 63.27 & 8.42\\
     % PointNovo &  &  &\\
    AdaNovo& 66.31  &67.92 &6.02\\
    \bottomrule
    \end{tabular}
    \label{cost_table}
\end{table}
\label{sec:appendix_cost}

\section{Sensitivity Analysis}
\begin{figure*}[ht]
    \centering
    \subfigure[Human (Peptide-level)]{
    \label{bar-pep}
    \includegraphics[width=0.4\textwidth]{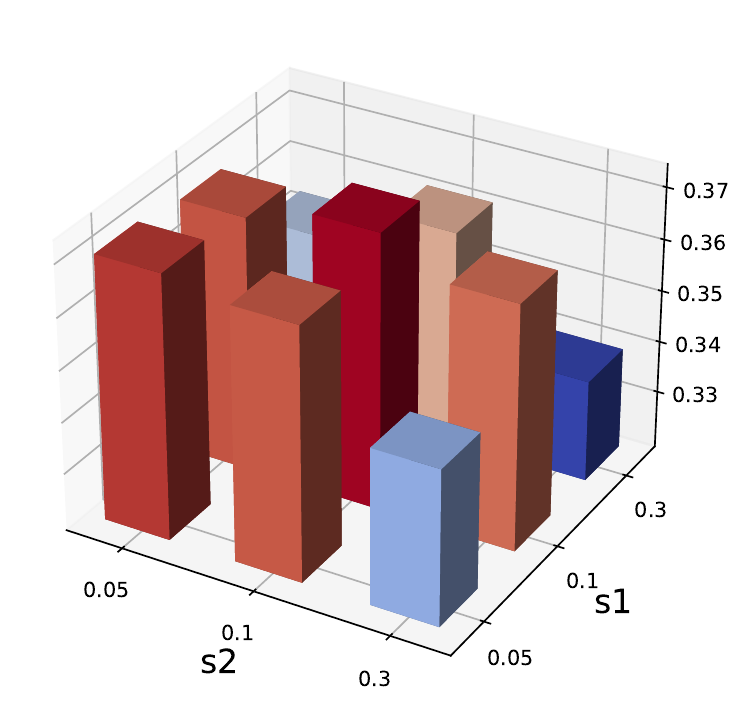}}
    \subfigure[Human (AA-level)]{
    \label{bar-aa}
    \includegraphics[width=0.4\textwidth]{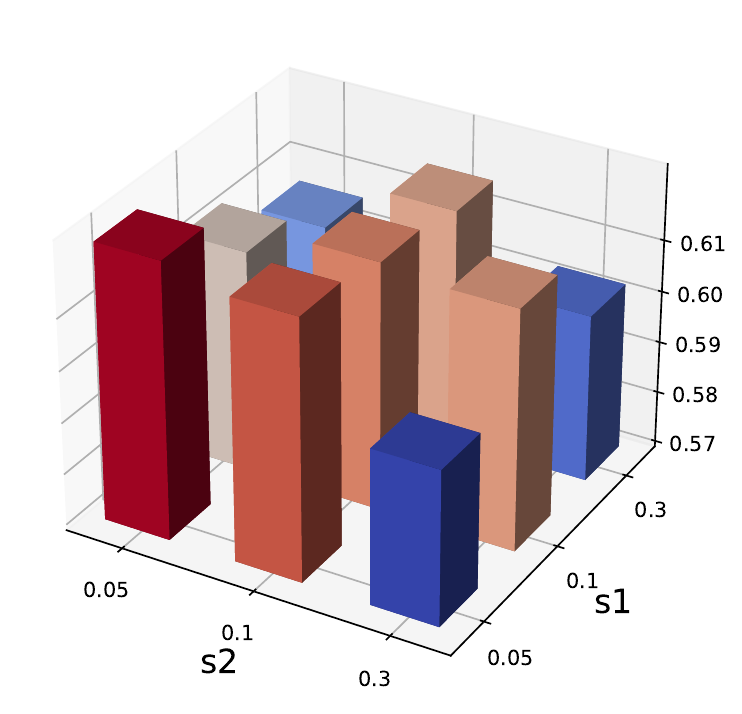}}
    \caption{The effects of the two hyperparameters $s_1$ and $s_2$ for adanovo. On the left are the peptide precision of the AdaNovo under different hyperparameter settings; on the right are the corresponding amino acid precision}
    \label{bar_pic}
\end{figure*}

In this section, we investigate the effects of the two hyperparameters $s_1$ and $s_2$, which determines the influence of amino acid-level and PSM-level training strategy. As shown in Figure~\ref{bar_pic}, we tune both $s_1$ and $s_2$ within the range $[0.05, 0.1, 0.3]$ and observe that the values of these two hyperparameters significantly affect the final performance of the model. Additionally, the optimal hyperparameters vary across different models, indicating differences in noise and the distributions of amino acids with PTMs  among different datasets. It is necessary to finely adjust the values of $s_1$ and $s_2$ based on the dataset, representing the balance between amino acid-level and PSM-level training strategies.
\label{sec:appendix_para}
% You can have as much text here as you want. The main body must be at most $8$ pages long.
% For the final version, one more page can be added.
% If you want, you can use an appendix like this one.  

% The $\mathtt{\backslash onecolumn}$ command above can be kept in place if you prefer a one-column appendix, or can be removed if you prefer a two-column appendix.  Apart from this possible change, the style (font size, spacing, margins, page numbering, etc.) should be kept the same as the main body.
%%%%%%%%%%%%%%%%%%%%%%%%%%%%%%%%%%%%%%%%%%%%%%%%%%%%%%%%%%%%%%%%%%%%%%%%%%%%%%%
%%%%%%%%%%%%%%%%%%%%%%%%%%%%%%%%%%%%%%%%%%%%%%%%%%%%%%%%%%%%%%%%%%%%%%%%%%%%%%%

\end{document}